\documentclass[twocolumn,10pt,final]{tsfp8}
\usepackage{flushend}
\usepackage[authoryear,round]{natbib}
\usepackage{units}
\usepackage{calc}
\usepackage[format=plain,justification=justified]{caption}

\newcount\ndots
\def\drawline#1#2{\raise 2.5pt\vbox{\hrule width #1pt height #2pt}}
\def\spacce#1{\hskip #1pt}
\def\solid{\drawline{24}{.5}\nobreak\ }
\def\bdash{\hbox{\drawline{4}{.5}\spacce{2}}}
\def\dashed{\bdash\bdash\bdash\bdash\nobreak\ }
\def\bdot{\hbox{\drawline{1}{.5}\spacce{2}}}
\def\dotted{\hbox{\leaders\bdot\hskip 24pt}\nobreak\ }

\def\chndot{\hbox {\drawline{9.5}{.5}\spacce{2}\drawline{1}{.5}\spacce{2}\drawline{9.5}{.5}}\nobreak\ }

\def\trian{\raise 1.25pt\hbox{$\scriptscriptstyle\triangle$}\nobreak\ }
\def\circle{$\circ$\nobreak\ }

\def\square{${\vcenter{\hrule height .4pt
        \hbox{\vrule width .4pt height 3pt \kern 3pt
        \vrule width .4pt}
        \hrule height .4pt}}$\nobreak\ }
\def\plus{\raise 1.25pt \hbox{$\scriptscriptstyle +$}\nobreak\ }
\def\x{\raise 1.25pt \hbox{$\scriptscriptstyle \times$}\nobreak\ }

\def\solidtrian{\raise 1.25pt
   \hbox to 3bp{
\hfill}\nobreak\ }
\def\solidsquare{\vrule height .9ex width .8ex depth -.1ex\nobreak\ }

\def\solidcclose{\drawline{10}{.5}\nobreak\raise
  0.5pt\hbox{$\bullet$}\drawline{10}{.5}\nobreak\ }

\def\solidsclose{\drawline{10}{.5}\nobreak\raise
  0.5pt\hbox{\solidsquare}\drawline{10}{.5}\nobreak\ }

\def\solidtclose{\drawline{10}{.5}\nobreak\raise
  0.5pt\hbox{\solidtrian}\drawline{10}{.5}\nobreak\ }

\def\solidcopen{\drawline{10}{.5}\nobreak\raise
  0.5pt\hbox{\circle}\drawline{10}{.5}\nobreak\ }

\def\solidsopen{\drawline{10}{.5}\nobreak\raise
  0.5pt\hbox{\square}\drawline{10}{.5}\nobreak\ }

\def\solidtopen{\drawline{10}{.5}\nobreak\raise
  0.5pt\hbox{\trian}\drawline{10}{.5}\nobreak\ }

\def\solidx{\drawline{10}{.5}\nobreak\raise
  0.5pt\hbox{\x}\drawline{10}{.5}\nobreak\ }

\usepackage{graphicx}
\usepackage{psfrag}
\usepackage{epstopdf}
\usepackage{amsmath,rotating}

\graphicspath{{figures/}}

\usepackage{url}

\usepackage{color}

\newcommand{\beq}{\begin{equation}}
\newcommand{\eeq}{\end{equation}}
\newcommand{\ol}[1]{\overline{#1}}
\newcommand{\wt}[1]{\widetilde{#1}}

\newcommand{\hk}[1]  {\begin{sideways}{#1}\end{sideways}}

\title{A parametrized non-equilibrium wall-model for large-eddy simulations}

\author{
  S. Hickel$^1$ ,
  E. Touber$^2$ ,
  J. Bodart$^3$ ,
  J. Larsson$^4$
  \affiliation{\small{$^1$ Faculty of Aerospace Engineering, TU Delft, NL.}\\
               \small{$^2$ Department of Mechanical Engineering, Imperial College, UK.}\\
               \small{$^3$ ISAE, Universit\'e de Toulouse, France.}\\
               \small{$^4$ Department of Mechanical Engineering, University of Maryland, USA}}}

\begin{document}

\setcounter{page}{1}

\maketitle

\fontsize{9}{11}\selectfont
\captionsetup{font=small}

\section*{ABSTRACT}

Wall-models are essential for enabling large-eddy
simulations (LESs) of realistic problems at high Reynolds numbers.
The present study is focused on approaches that directly model the
wall shear stress, specifically on filling the gap between models
based on wall-normal ordinary differential equations (ODEs) 
that assume equilibrium and models based on
full partial differential equations (PDEs) that do not.
We develop ideas for how to incorporate non-equilibrium effects (most
importantly, strong pressure-gradient effects) in the wall-model while
still solving only wall-normal ODEs. We test these ideas using two
reference databases: an adverse
pressure-gradient turbulent boundary-layer and a shock/boundary-layer
interaction problem, both of which lead to separation
and re-attachment of the turbulent boundary layer.


\section*{\uppercase{Introduction}}

\begin{figure*}[t!]
\hspace{5em}{\small{ \textbf{a)}\ \ Test case 1 : STBLI}}
  \begin{center}
    \begin{minipage}[t]{0.75\textwidth}
      \includegraphics[width=\textwidth,clip=]{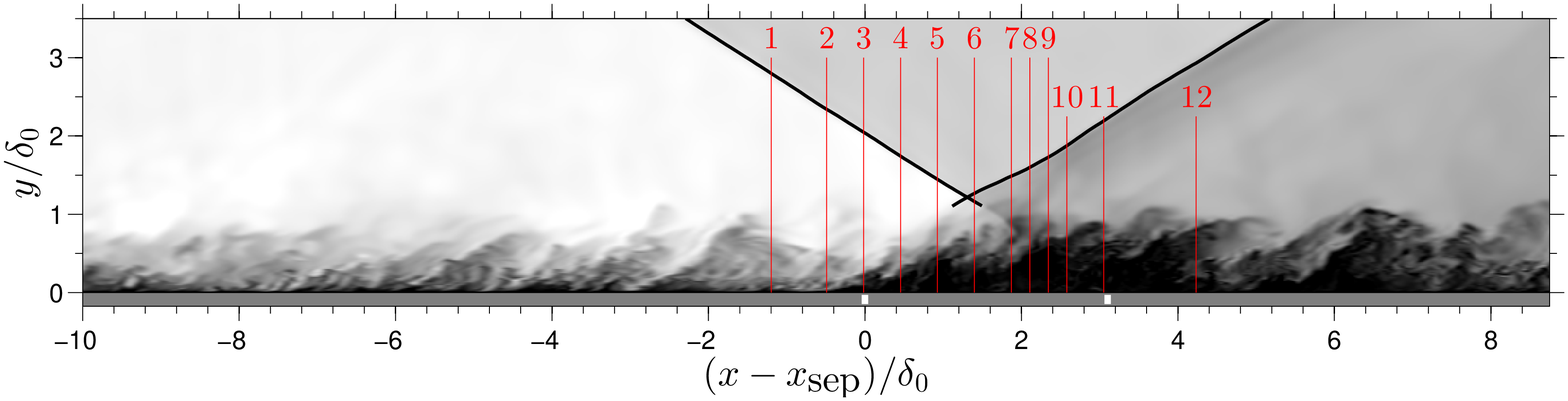}
    \end{minipage}
  \end{center}
\hspace{5em}{\small{ \textbf{b)}\ \ Test case 2 : APGTBL}}
  \begin{center}    
    \begin{minipage}[t]{0.75\textwidth}
      \includegraphics[width=\textwidth,clip=]{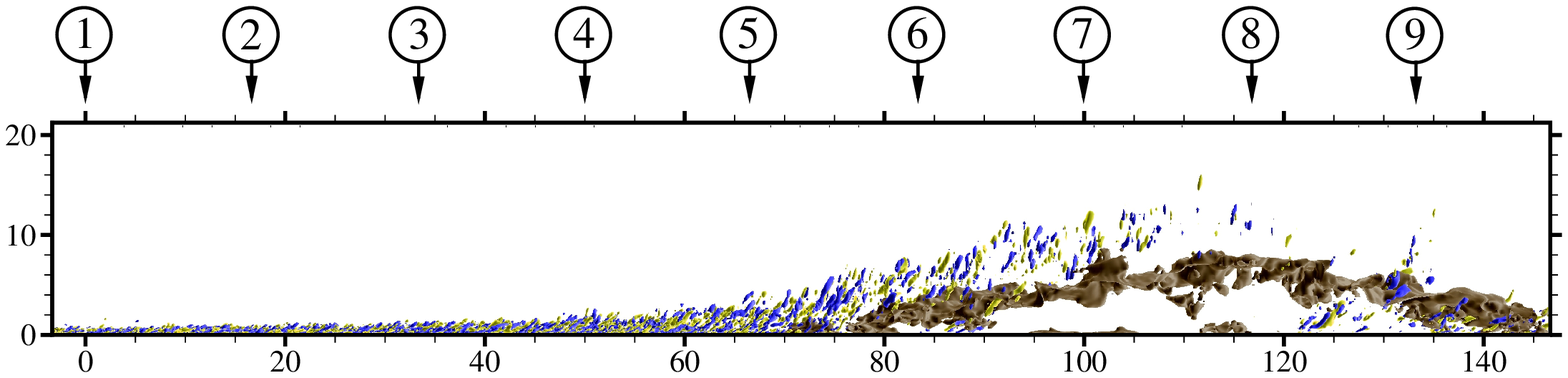}
      \put( -335.0, 42.0 ){\makebox(0,0)[r]{\footnotesize\hk{$y / \delta_0$}}}
      \put(  -30.0, -3.0 ){\makebox(0,0)[r]{\footnotesize$x / \delta_0$}}
      \vspace{-1em}
    \end{minipage}    
  \end{center}
  \captionsetup{singlelinecheck=off}
  \caption[Test cases and interrogation stations used in this study.]{        \label{fig_overview} 
      Test cases and interrogation stations used in this study.
      \begin{enumerate}
      \item[a)]{The 8-degree IUSTI shock/turbulent-boundary-layer interaction
      (labeled STBLI throughout) experiment of \cite{dupont:06}, with wall-resolved reference LES by \cite{touber:09}.
      Visualization of the instantaneous temperature.}
      \item[b)]{The adverse pressure-gradient turbulent
      boundary-layer (labeled APGTBL throughout) studied in \cite{hickel:08}.
      Visualization of the instantaneous coherent structures (Q-criterion) and the
      separation bubble($u_1=0$ iso-surfaces).}
      \end{enumerate}
      }
  \captionsetup{singlelinecheck=on}
\end{figure*}

Large-eddy simulations (LES) have become part of the basic toolkit for
fundamental fluids research.
However, the ``near-wall problem'' of requiring essentially DNS-type
grid resolution in the innermost layer of turbulent boundary layers
has effectively prevented LES from being applied to many realistic
turbulent flows
\cite[cf.][]{piomelli:02}.
The solution is to model (rather than resolve) the inner part of
turbulent boundary layers, say the innermost 10-20\% of the boundary
layer thickness $\delta$.
By doing this, the grid resolution in an LES is set solely by the need
to resolve the remaining outer layer.

There are essentially two different classes of methods that follow
this approach.
In hybrid LES/RANS and detached eddy simulation (DES), the unsteady
evolution equations with an eddy viscosity term are solved everywhere
in the domain.
The eddy viscosity is then taken from some RANS-type model in
the inner layer and some LES-type subgrid model in the outer layer
everywhere else in the flow.
A second class of methods instead models the wall-stress directly.
The LES domain is then defined formally as extending all the way to
the wall, while an auxiliary set of equations is solved in an
overlapping layer covering the innermost 10-20\% of $\delta$.
These auxiliary equations are forced by the LES at their upper
boundary, and feed the computed wall shear stress and heat
transfer back to the LES. The focus in this study is exclusively on the
second class of methods.\\

The most obvious models for the wall stress assume equilibrium; that is,
they neglect both the convective and pressure-gradient terms in
addition to the wall-parallel diffusive terms.
For constant-density flows above the viscous layer, these models yield
the famous log-law.
While the wall-model only models the innermost 10-20\% of $\delta$
(and the LES directly resolves the outermost 80-90\%),
there has been a long interest in removing the assumption of
equilibrium from the wall-model, in hopes of making wall-modeled LES
capable of more accurate predictions in the presence of flow
separation, etc.

One approach to including non-equilibrium effects was pioneered by
\cite{balaras:96}, who solved the thin boundary layer equations
(including convection and pressure-gradient, neglecting only
wall-parallel diffusion) as a wall-model.
From a practical point-of-view, the main drawback of this approach is
that a partial differential equation (PDE) must be solved as the
wall-model. Thus a full grid with neighbor connectivity is needed
in the near-wall layer, in addition to the already existing LES grid.
This is a serious obstacle if one seeks to implement the wall-model in
an unstructured code for complex geometries.
In fact, one could argue that any new wall-model (of the wall-stress
kind) will be broadly adopted only if it involves at most connectivity
(i.e., derivatives) in the wall-normal direction and time, both of
which are easily implemented in a general unstructured code framework.

The challenge, therefore, is to include non-equilibrium effects
without the need for wall-parallel derivatives.
\cite{hoffmann:95} and \cite{chen:13} included the pressure-gradient and the
temporal term but excluded the convective term\footnote{Throughout this
  paper, \emph{temporal term} refers to $\partial_t u_i$ while
         \emph{convective term} refers to $u_j \partial_j u_i$.}.
Since the pressure-gradient is constant throughout the wall-modeled
layer, and imposed from the LES, this approach does not require
wall-parallel derivatives within the wall-model.
\cite{wang:02} and subsequently \cite{catalano:03} went one step
further and retained only the pressure-gradient term in calculations
of the flow over a trailing edge and a circular cylinder,
respectively.
Neither of these approaches is satisfactory, for reasons to be shown below.

The objective of the present study is to develop a wall-model that
includes non-equilibrium effects while still requiring only numerical
connectivity in, at most, the wall-normal and temporal directions.
\newpage
\noindent
Towards this end, the present paper will:
\begin{enumerate}
\item
Argue and show that past attempts at including or neglecting the
temporal, convective and pressure-gradient terms independently are
inconsistent, in the sense that the temporal and convective terms
jointly describe the evolution of a fluid particle and that the
pressure-gradient and convective terms largely balance outside of the
viscous sublayer.
\item
Argue and show that the convective term can be parametrized in
terms of outer layer LES quantities, thereby eliminating the need
for wall-parallel derivatives in the wall-model.
\end{enumerate}
%


\section*{\uppercase{Time-filtered equations}}

When implemented in an LES code, the wall-model is continuously forced
by the LES at the upper boundary of the wall-modeled domain (say, at
height $h_{\rm wm}$).
To give accurate results, $h_{\rm wm}$ should be within the inner
part of the boundary layer, so about 10-20\% of the boundary layer
thickness $\delta$ or less.
For accuracy, the grid spacing in the LES needs to be sufficiently
small compared to $h_{\rm wm}$
\cite[][]{kawai:12:wallmodeling},
which implies that $h_{\rm wm}$ should not be chosen too small.

The continuous forcing by the LES at the top boundary means that the
wall-model operates in an unsteady mode.
The relatively large (RANS-type) eddy viscosity in the wall-model
acts as a low-pass filter; therefore, the solution in the wall-modeled layer
will be unsteady with primarily low frequencies.
This is approximately accounted for in the analysis below by
applying a low-pass filter to the wall-resolved LES databases,
specifically a top-hat filter with characteristic width $\tau$ defined as
\beq
\ol{u}(t;\tau) = \frac{1}{\tau} \mathop{\int}_{-\tau/2}^{\tau/2} u(t-t')
dt'
\,.
\eeq
With density-weighting, the associated Favre filter is $\wt{u} =
\ol{\rho u} / \ol{\rho}$.
Application of this filter to the streamwise momentum equation yields to leading order
\beq
\label{eqn_timefiltered_momentum}
\partial_t \wt{u}
+
\wt{u}_j \partial_j \wt{u}
-
\frac{
\partial_2 \left[ \mu(\ol{T}) \, \partial_2 \wt{u} \right]
}{\ol{\rho}}
\approx
-
\frac{
\partial_1 \ol{p}
}{\ol{\rho}}
-
\underbrace{
\frac{
\partial_j (\ol{\rho} \left[ \wt{u_ju} - \wt{u}_j \wt{u} \right])
}{\ol{\rho}}
}_{\rm modeled}
\,,
\eeq
where wall-parallel diffusion has been neglected, as well as terms due
to nonlinearity in the temperature-dependence of the viscosity.
The streamwise, wall-normal and spanwise directions
(perhaps defined locally) are denoted by subscripts 1, 2 and 3,
respectively.
For brevity, the streamwise velocity is interchangeably labeled $u$ or
$u_1$.

\begin{figure*}[t]
  \begin{center}

    \begin{minipage}[t]{0.4\textwidth}
      \includegraphics[width=\textwidth,height=0.8\textwidth,clip=]{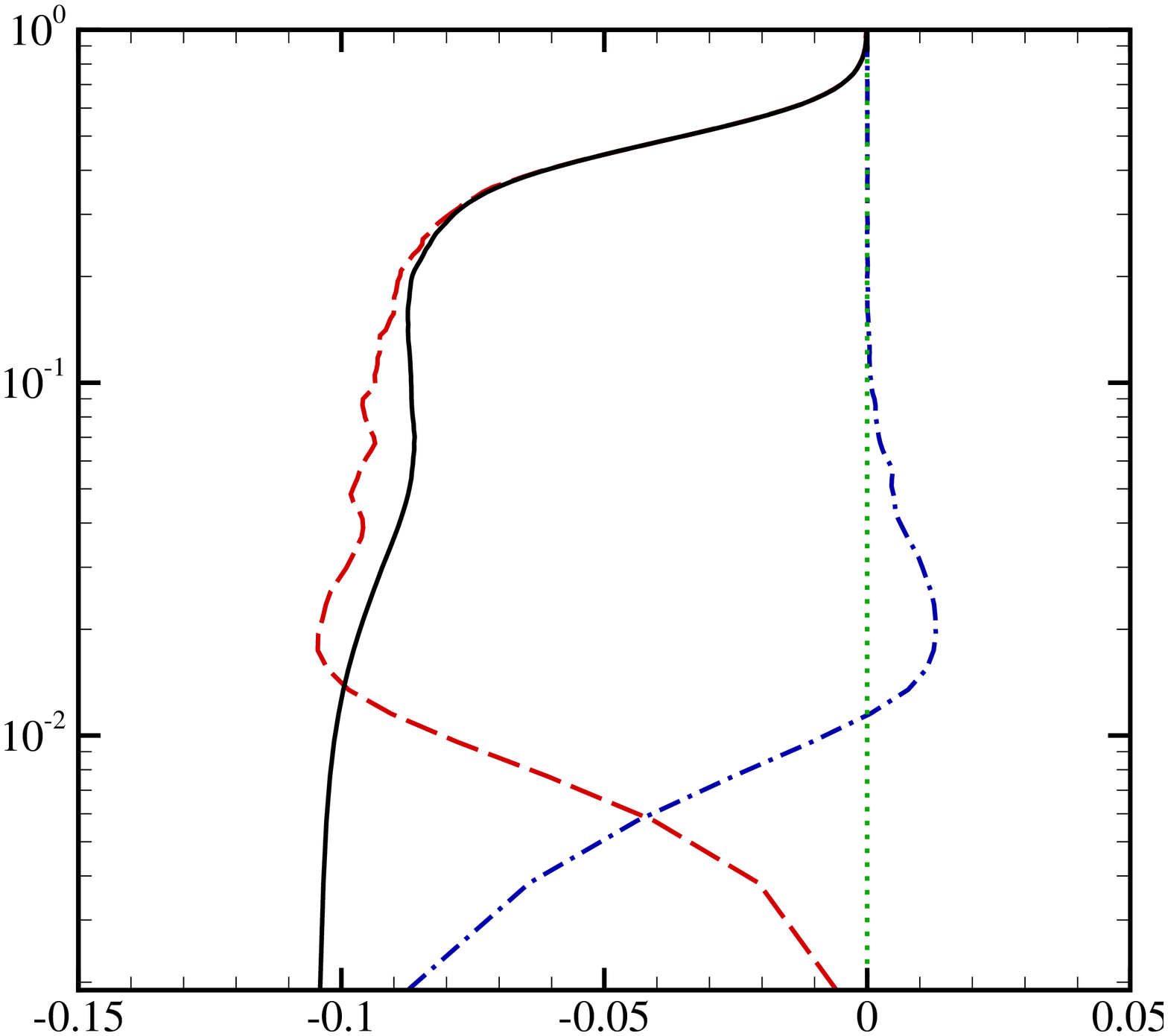}
      \put( -170.0,130.0 ){\makebox(0,0)[r]{\footnotesize$ y / \delta_0 $}}
      \put(  -80.0,  5.0 ){\makebox(0,0)[c]{\footnotesize time averaged budget}}
    \end{minipage}
    \hspace{2em}
    \begin{minipage}[t]{0.4\textwidth}
      \includegraphics[width=\textwidth,height=0.8\textwidth,clip=]{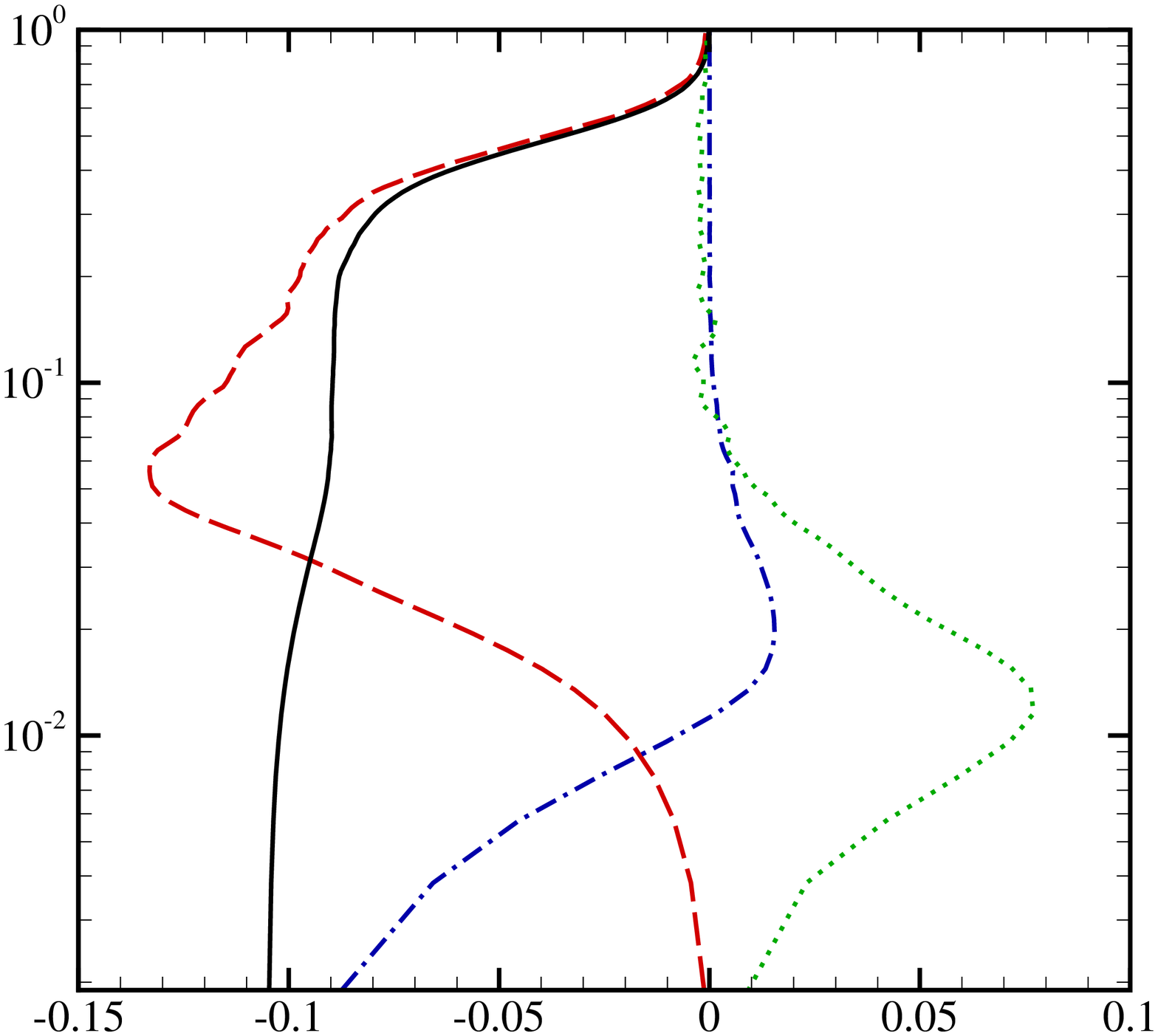}
      \put( -170.0,130.0 ){\makebox(0,0)[r]{\footnotesize$ y / \delta_0 $}}
      \put(  -80.0,  5.0 ){\makebox(0,0)[c]{\footnotesize time averaged budget}}
    \end{minipage}
    \begin{minipage}[t]{0.4\textwidth}
      \includegraphics[width=\textwidth,height=0.8\textwidth,clip=]{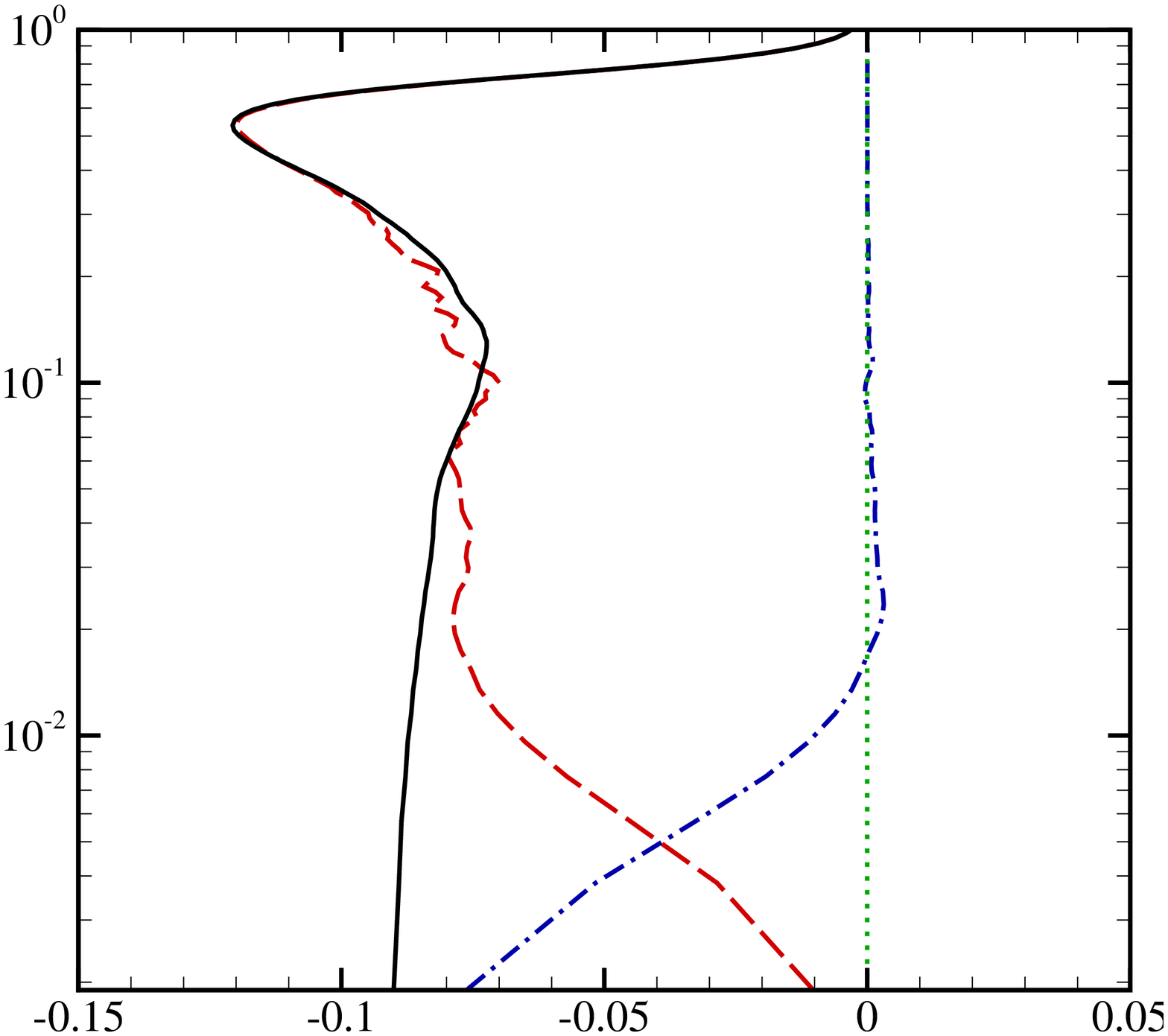}
      \put( -170.0,130.0 ){\makebox(0,0)[r]{\footnotesize$ y / \delta_0 $}}
      \put(  -80.0,  5.0 ){\makebox(0,0)[c]{\footnotesize time averaged budget}}
    \end{minipage}
    \hspace{2em}
    \begin{minipage}[t]{0.4\textwidth}
      \includegraphics[width=\textwidth,height=0.8\textwidth,clip=]{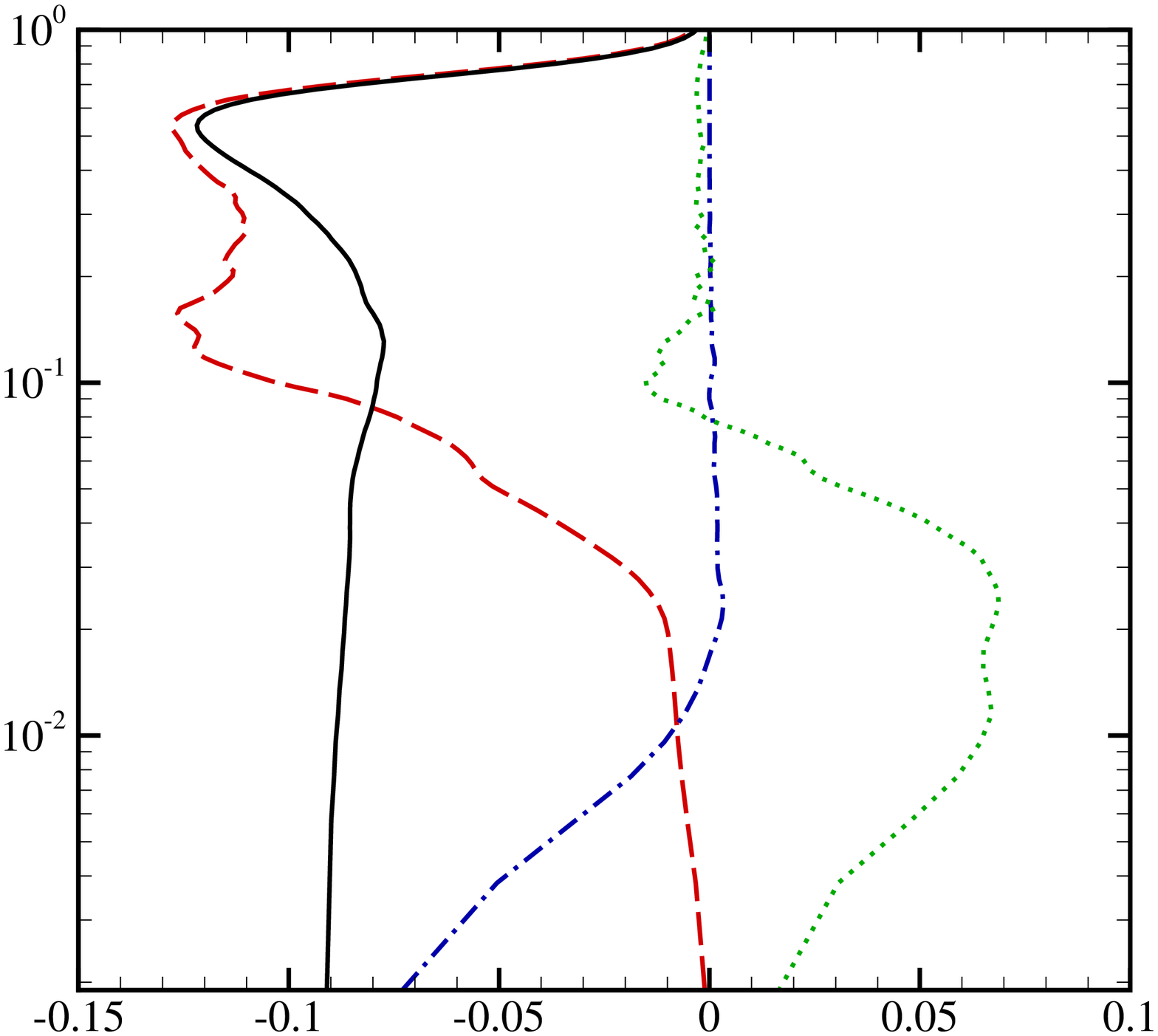}
      \put( -170.0,130.0 ){\makebox(0,0)[r]{\footnotesize$ y / \delta_0 $}}
      \put(  -80.0,  5.0 ){\makebox(0,0)[c]{\footnotesize time averaged budget}}
    \end{minipage}
    \caption{ \label{fig_momentum_balance_emile}
      Averaged terms in Eq.~(\ref{eqn_timefiltered_momentum}) for the
      STBLI case at stations 3 (upper
      row) and 4 (lower row),
      filtered using time-scales $ \tau U_\infty / \delta \approx  0.0 $ (left column) 
      and $ \tau U_\infty / \delta \approx  8.5 $ (right column).
      Pressure-gradient (\solid) ,
       viscous term (\chndot),
      resolved convection in all three directions (\dashed),
      and
      unresolved wall-normal Reynolds stress  (\dotted).
    }
  \end{center}
\end{figure*}

%

\section*{\uppercase{Test cases}}

As a first step in this study, data from two reference large-eddy
simulations is used to assess the new ideas in an {a priori}
manner.
The two reference LESs used sufficiently fine grids to fully resolve
the viscous near-wall layer in a quasi-DNS sense.

The first case considered is wall-resolving LES 
\citep{touber:09} of a shock/turbulent-boundary-layer
interaction (labeled STBLI throughout) consistent with the flow conditions of the IUSTI
experiment of \cite{dupont:06}.
An oblique shock wave generated by an 8-degree wedge 
impinges on a $Ma=2.3$ flat-plate turbulent-boundary layer with a
displacement-thickness Reynolds number of $Re_{\delta1} = 21000$.
The test case provides regions where equilibrium assumptions are
supposed to hold and regions with strong non-equilibrium effects.
A visualization of this flow is shown in Figure~\ref{fig_overview}a)
based on the instantaneous temperature for the LES of
\cite{touber:09}.

The second test case is the incompressible non-equilibrium turbulent
flat-plate boundary-layer flow of \cite{hickel:08} with a
displacement-thickness Reynolds number going from $Re_{\delta1} = 1000$
to 30000.
Due to the strong non-equilibrium conditions, which result from a constant
adverse pressure-gradient imposed at the upper domain boundary,
the mean velocity profiles of this boundary layer flow do not follow
the classic logarithmic law of the wall.
The adverse pressure-gradient decelerates the flow and eventually leads to a highly unsteady
and massive flow separation,
which is not fixed in space and covers more than a third of the
computational domain.
The separated flow region and instantaneous coherent structures are
visualized in Figure~\ref{fig_overview}b) through an instantaneous
iso-surface of $u_1 = 0$ and iso-surfaces of the Q-criterion,
respectively.
This case is labeled APGTBL throughout this paper.

\section*{\uppercase{Consistency with the \newline Bernoulli equation}}
\label{section_consistency}

A typical wall-model essentially solves Eq.~(\ref{eqn_timefiltered_momentum})
with the unresolved convective term parametrized using an eddy
viscosity model.
The by far most common approach is to assume equilibrium, i.e., to neglect
the temporal, resolved convective and pressure-gradient terms.
This is exact only for Couette flow, but is a good
approximation in many cases.

\cite{hoffmann:95} and, later on, two studies coming out of the Center for
Turbulence Research
\cite[][]{wang:02,catalano:03}
retained the pressure-gradient term in Eq.~(\ref{eqn_timefiltered_momentum})
but neglected the convective term.
One objective of this paper is to point out that this is inconsistent.
Consider a flow with a non-zero pressure gradient.
In the limit of weak turbulence, for flow in a straight line
sufficiently far from the wall such that viscous effects are negligible, 
Eq.~(\ref{eqn_timefiltered_momentum}) should reduce to the so-called ``Euler-s''
equation, or, more familiarly (after integration along a streamline), to the Bernoulli
equation.
In other words, a non-zero pressure-gradient is accompanied by
accelerating/decelerating flow, which causes a non-zero streamwise
convective term.
Therefore, if the pressure-gradient term is explicitly included in the
wall-model, then the streamwise convective term must also be included
to satisfy this minimal consistency requirement.

\begin{figure*}[t]
  \begin{center}
    \begin{minipage}[t]{0.4\textwidth}
      \includegraphics[width=\textwidth,height=0.8\textwidth]{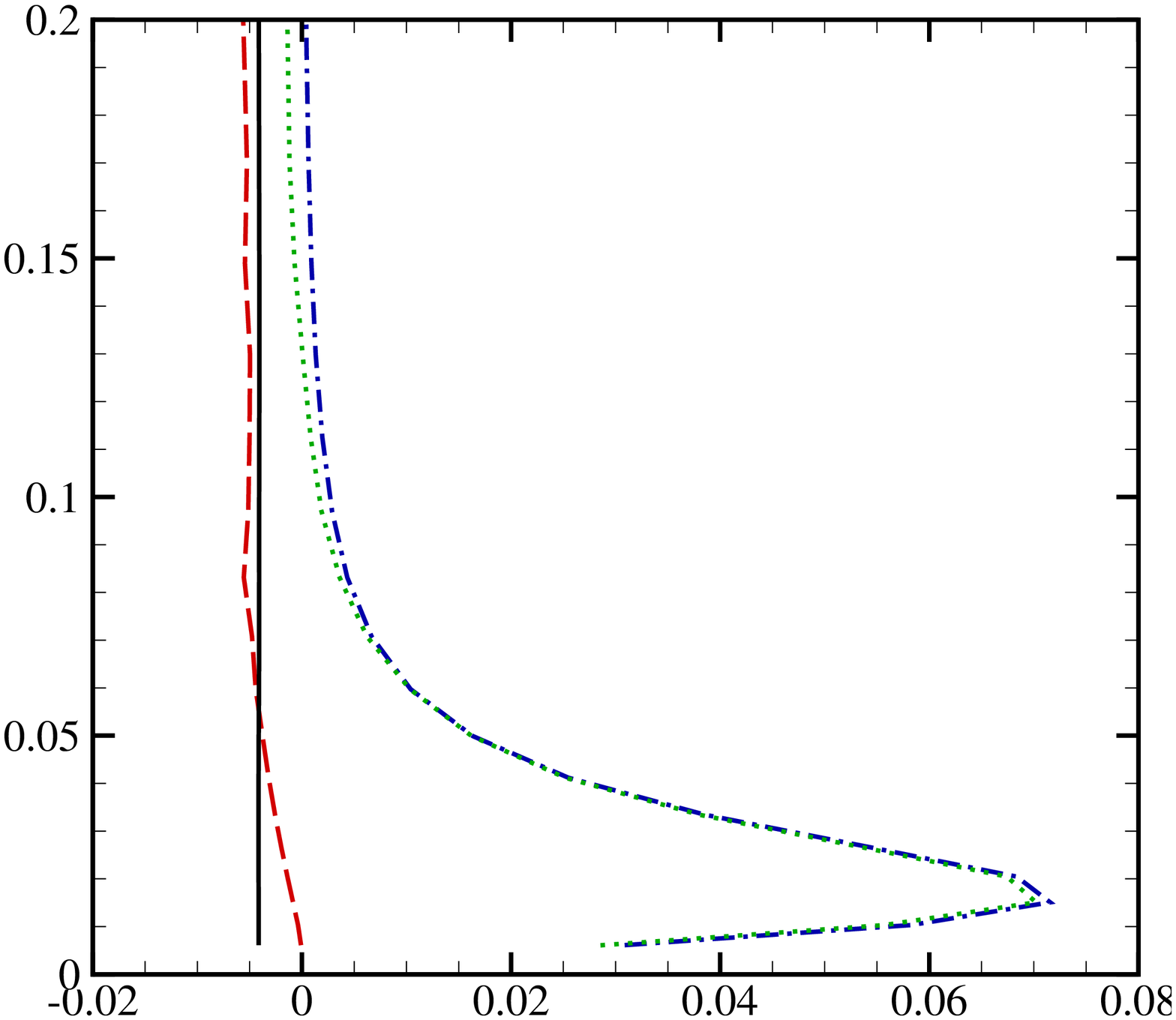}
      \put( -165.0, 130.0 ){\makebox(0,0)[r]{\footnotesize$y / \delta $}}
      \put(  -80.0,   5.0 ){\makebox(0,0)[c]{\footnotesize time averaged budget}}
    \end{minipage}
    \hspace{2em}
    \begin{minipage}[t]{0.4\textwidth}
      \includegraphics[width=\textwidth,height=0.8\textwidth]{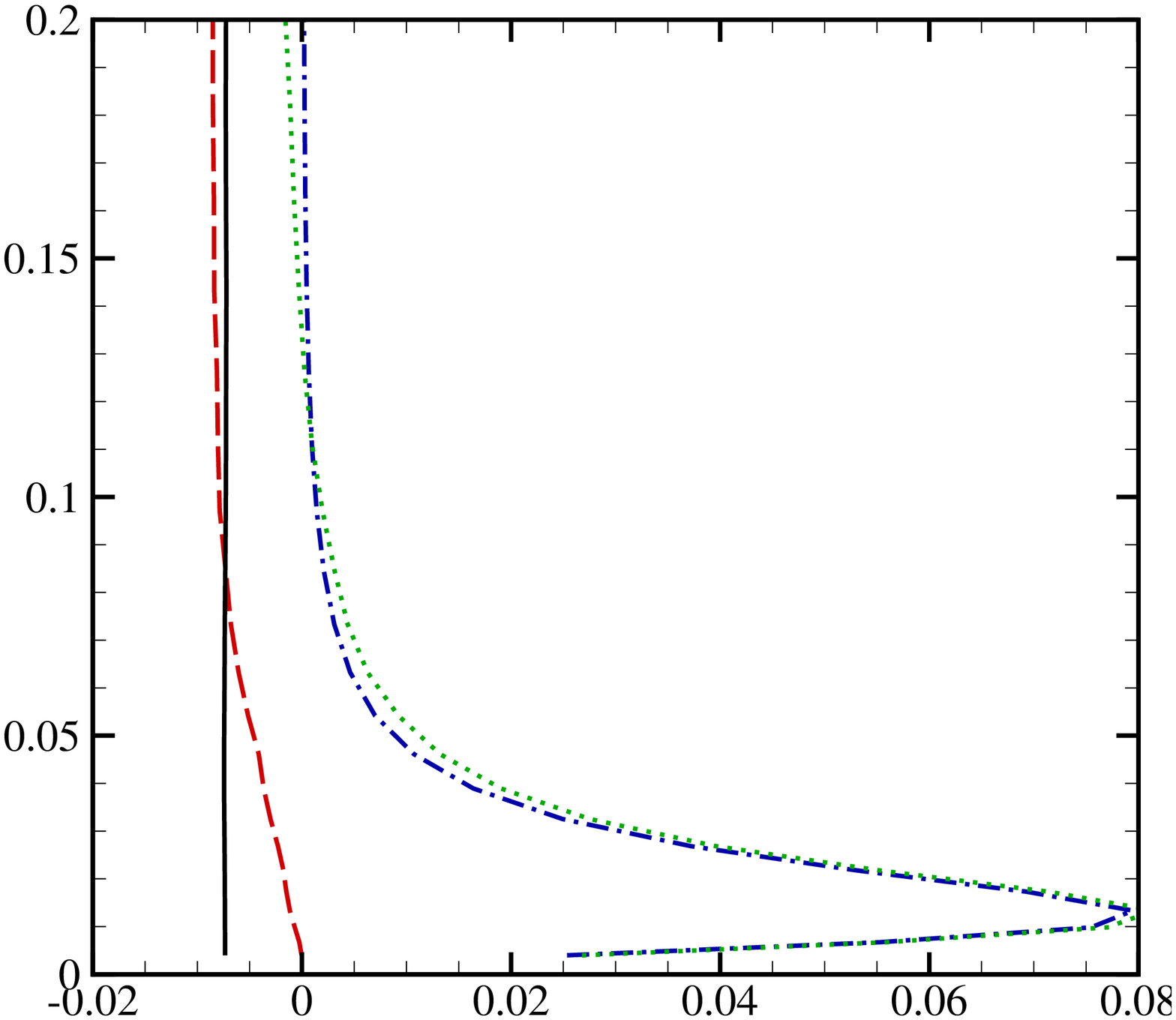}
      \put( -165.0,130.0 ){\makebox(0,0)[r]{\footnotesize$ y / \delta $}}
      \put(  -80.0,  5.0 ){\makebox(0,0)[c]{\footnotesize time averaged budget}}
      \vspace{1em}
    \end{minipage}
    \begin{minipage}[t]{0.4\textwidth}
      \includegraphics[width=\textwidth,height=0.8\textwidth]{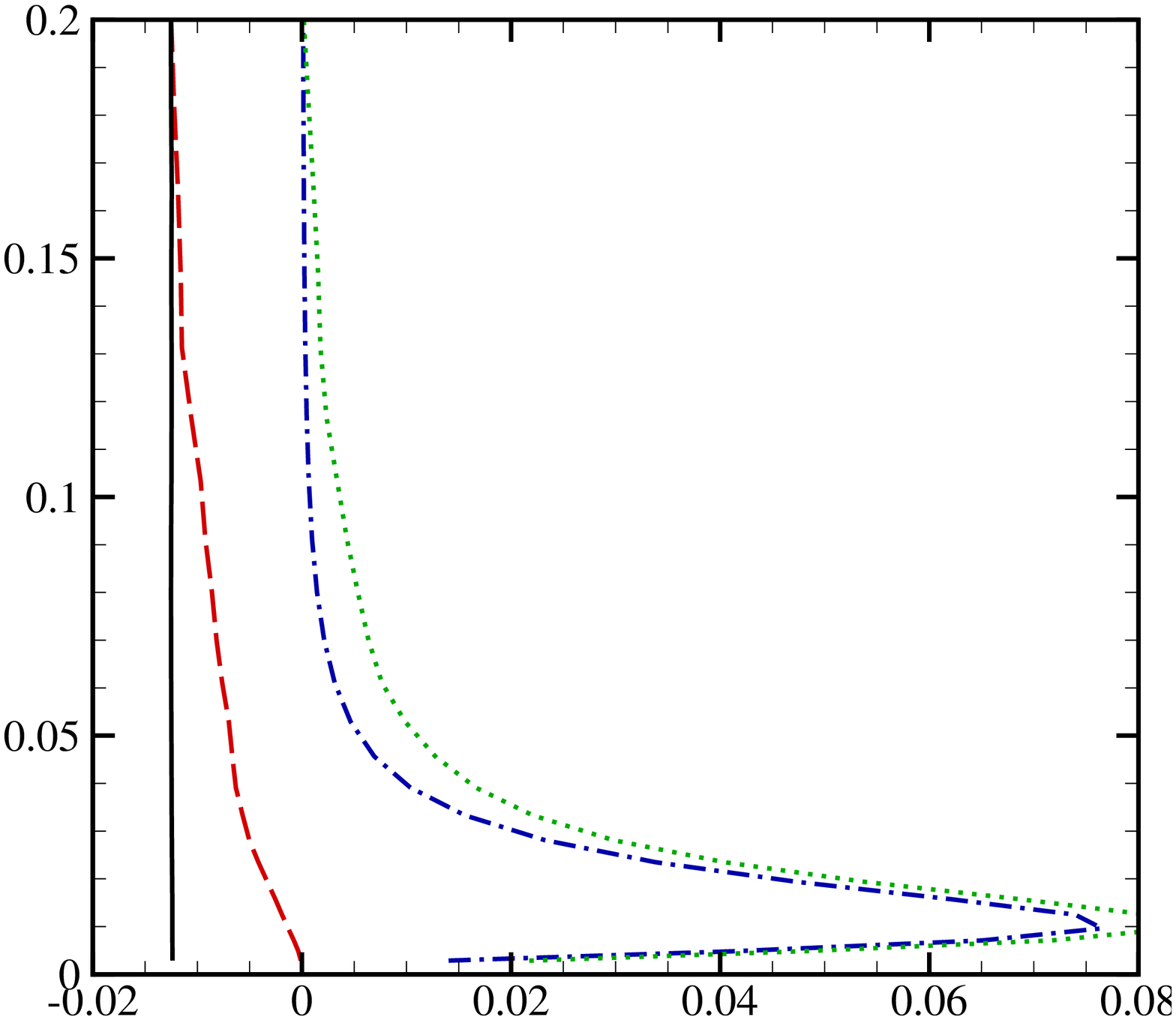}
      \put( -165.0, 130.0 ){\makebox(0,0)[r]{\footnotesize$y / \delta $}}
      \put(  -80.0,   5.0 ){\makebox(0,0)[c]{\footnotesize time averaged budget}}
    \end{minipage}
    \hspace{2em}
    \begin{minipage}[t]{0.4\textwidth}
      \includegraphics[width=\textwidth,height=0.8\textwidth]{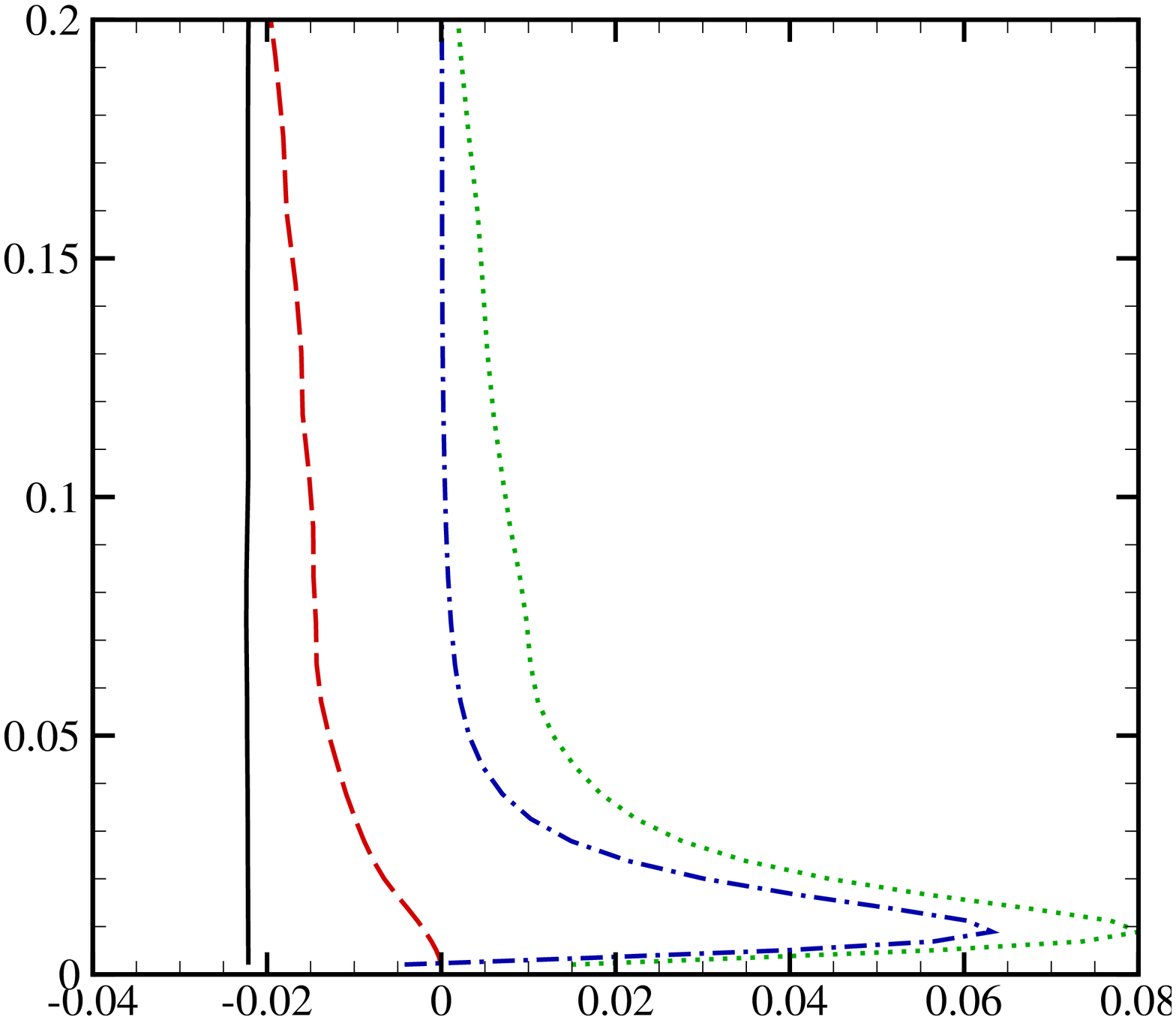}
      \put( -165.0,130.0 ){\makebox(0,0)[r]{\footnotesize$ y / \delta $}}
      \put(  -80.0,  5.0 ){\makebox(0,0)[c]{\footnotesize time averaged budget}}
    \end{minipage}
    \caption{ \label{fig_momentum_balance_stefan}
      Averaged terms in Eq.~(\ref{eqn_timefiltered_momentum}) for the
      APGTBL case
      at stations 1 to 4 using the time-scale $\tau \rightarrow \infty$.
      Pressure-gradient (\solid) ,
       viscous term (\chndot),
      resolved convection in all three directions (\dashed),
      and
      unresolved wall-normal Reynolds stress  (\dotted).
      All quantities are given in outer scaling, i.e., scaled by the local boundary-layer edge velocity.
    }
  \end{center}
\end{figure*}

Evidence of this is shown in Figures~\ref{fig_momentum_balance_emile}
and \ref{fig_momentum_balance_stefan}, which
show selected terms in the streamwise momentum equation
(\ref{eqn_timefiltered_momentum}) for the two test cases.
Results are shown both unfiltered and filtered in time for the first
test case, roughly
mimicking the effect of the RANS-type eddy viscosity.
The filtering hardly affects the viscous and pressure-gradient terms
at all, which is to be expected given their essentially linear
nature.
Note also that not all terms are shown in the figures; thus the sum of
all lines is not exactly zero.

The pressure-gradient is essentially balanced by the time-filtered
convection term (i.e., the convection that can be resolved by a wall-model)
in most of the outer part of the boundary layer, and only within the
viscous region does this approximate balance between convection and
pressure-gradient break down.

%

\section*{\uppercase{Convective parametrization}}
\label{section_parametrization}

%
\begin{figure*}
  \begin{center}
    \begin{minipage}[t]{0.4\textwidth}
      \includegraphics[width=\textwidth,height=0.8\textwidth]{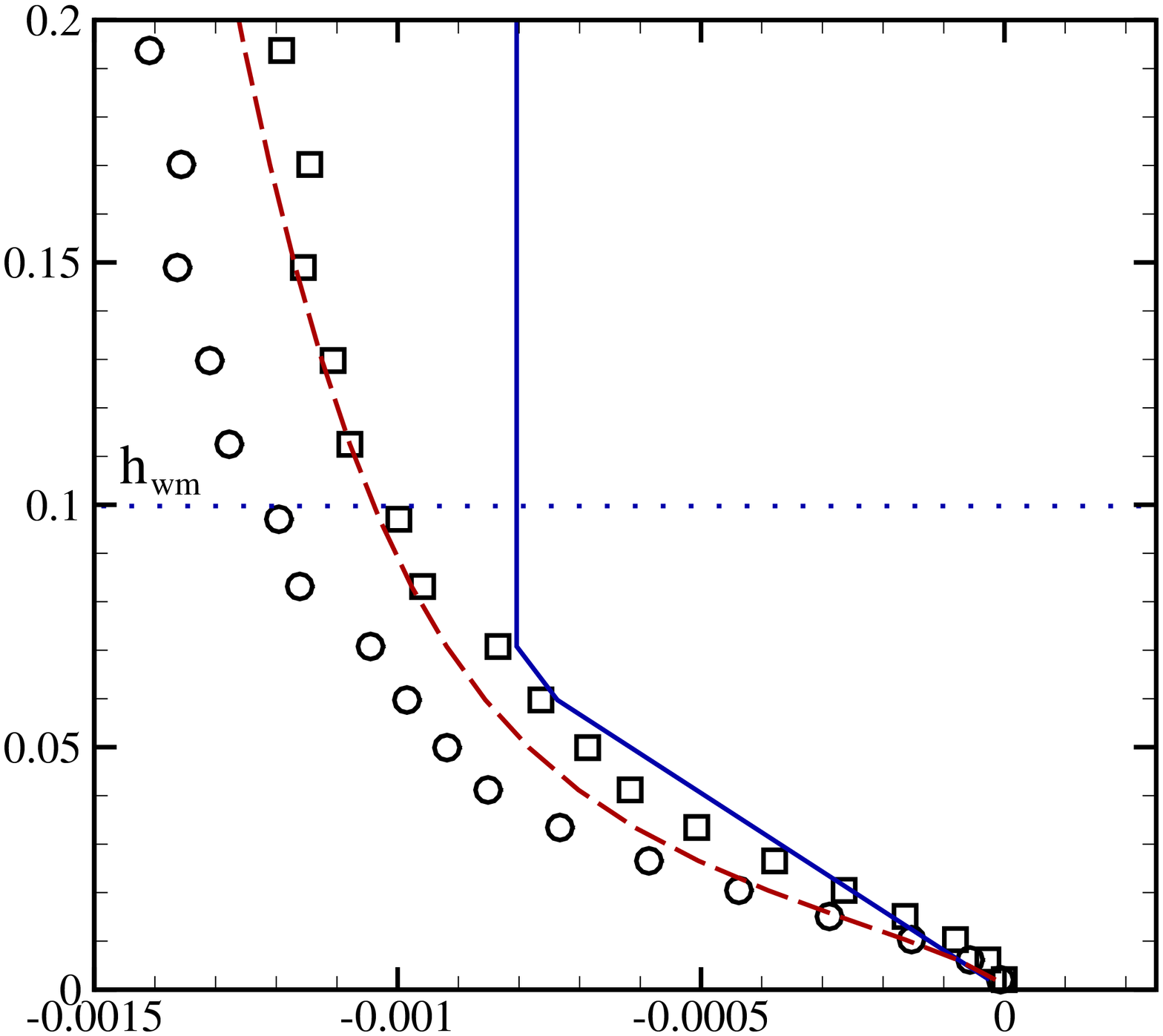}
      \put( -165.0, 130.0 ){\makebox(0,0)[r]{\footnotesize$y / \delta $}}
      \put(  -80.0,   5.0 ){\makebox(0,0)[c]{\  }}
    \end{minipage}
    \hspace{2em}
    \begin{minipage}[t]{0.4\textwidth}
      \includegraphics[width=\textwidth,height=0.8\textwidth]{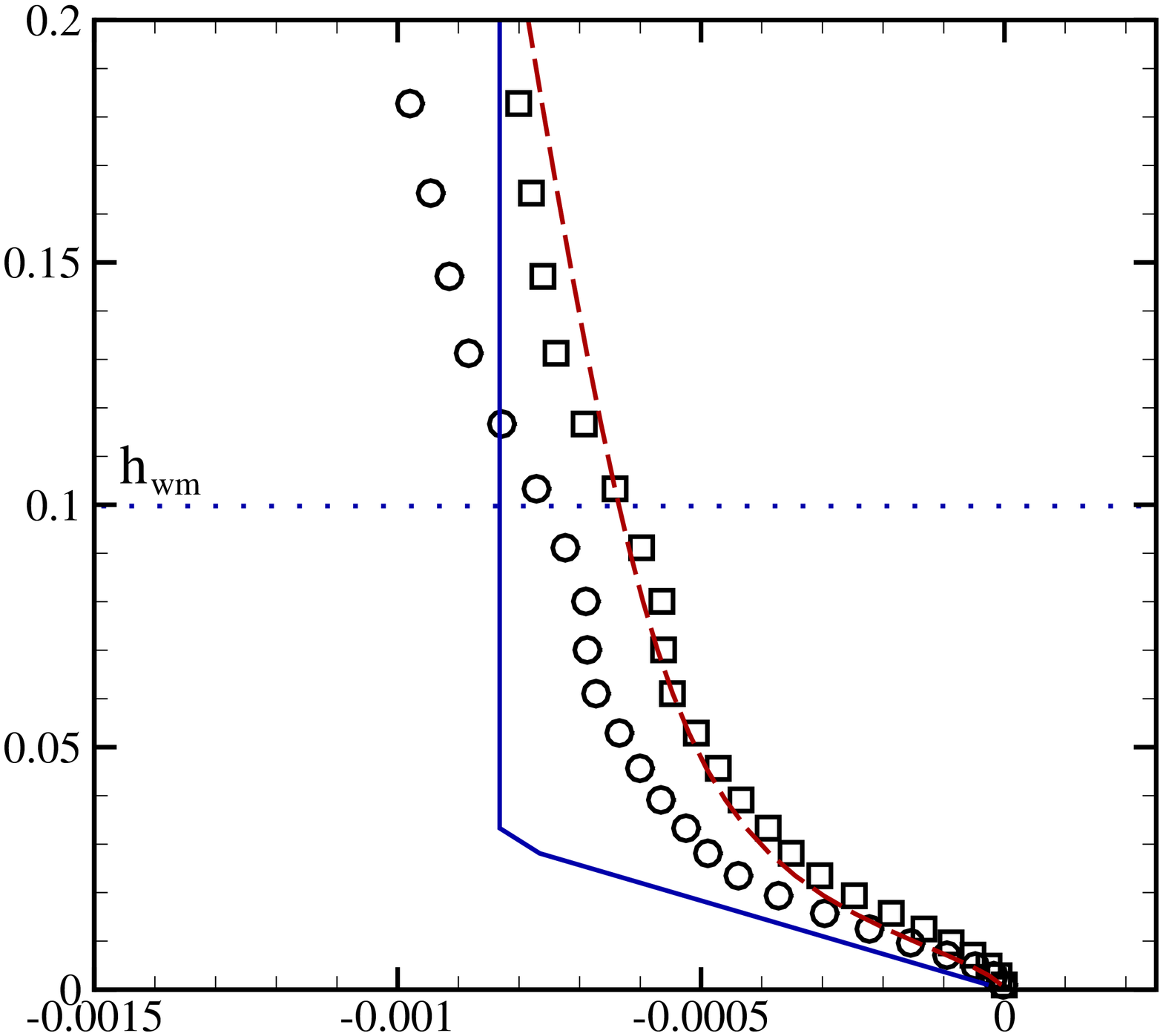}
      \put( -165.0,130.0 ){\makebox(0,0)[r]{\footnotesize$ y / \delta $}}
      \put(  -80.0,  5.0 ){\makebox(0,0)[c]{\  }}
    \end{minipage}
    \begin{minipage}[t]{0.4\textwidth}
      \includegraphics[width=\textwidth,height=0.8\textwidth]{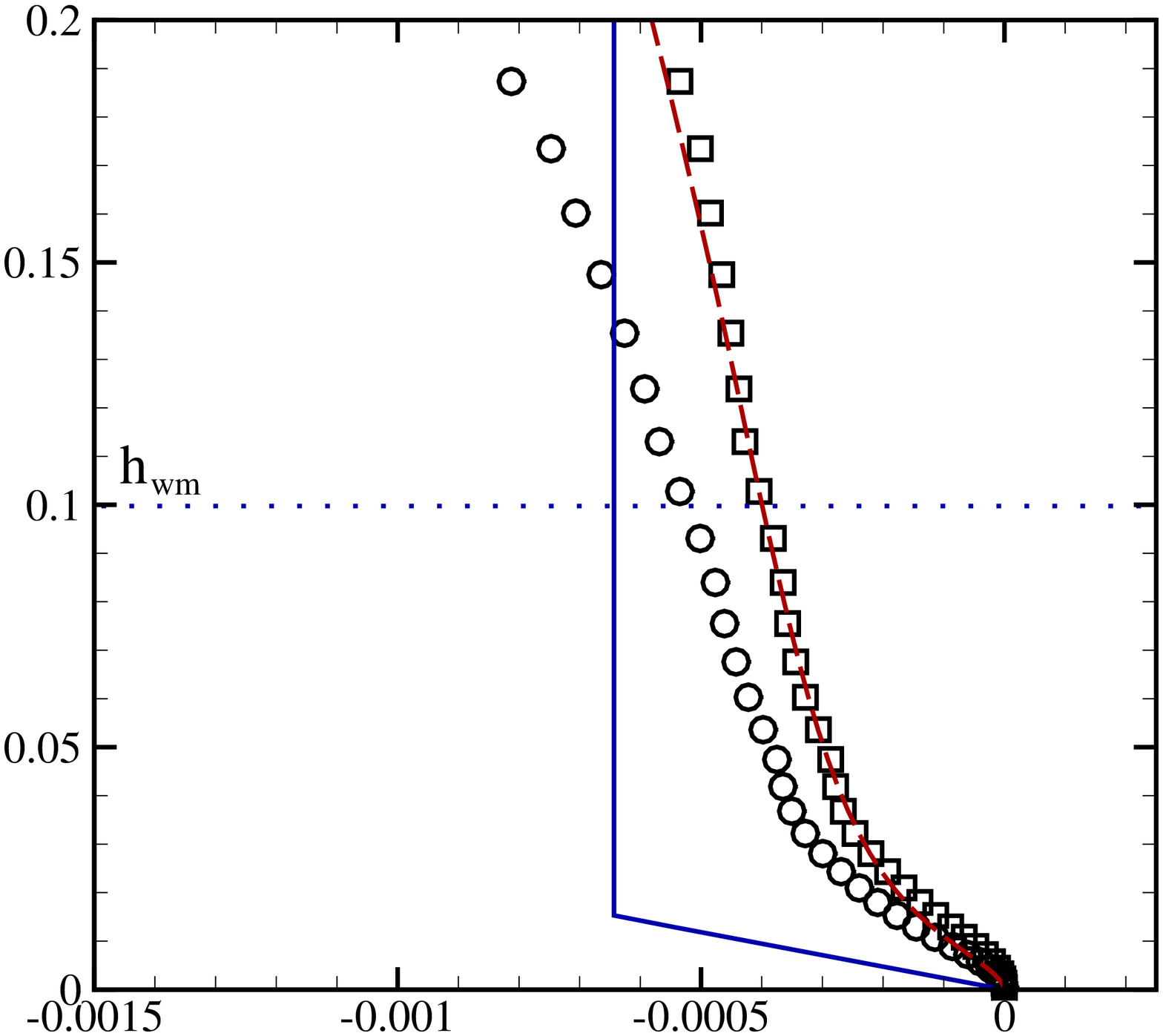}
      \put( -165.0,130.0 ){\makebox(0,0)[r]{\footnotesize$y / \delta $}}
      \put(  -80.0,  5.0 ){\makebox(0,0)[c]{\  }}
    \end{minipage}
    \hspace{2em}
    \begin{minipage}[t]{0.4\textwidth}
      \includegraphics[width=\textwidth,height=0.8\textwidth]{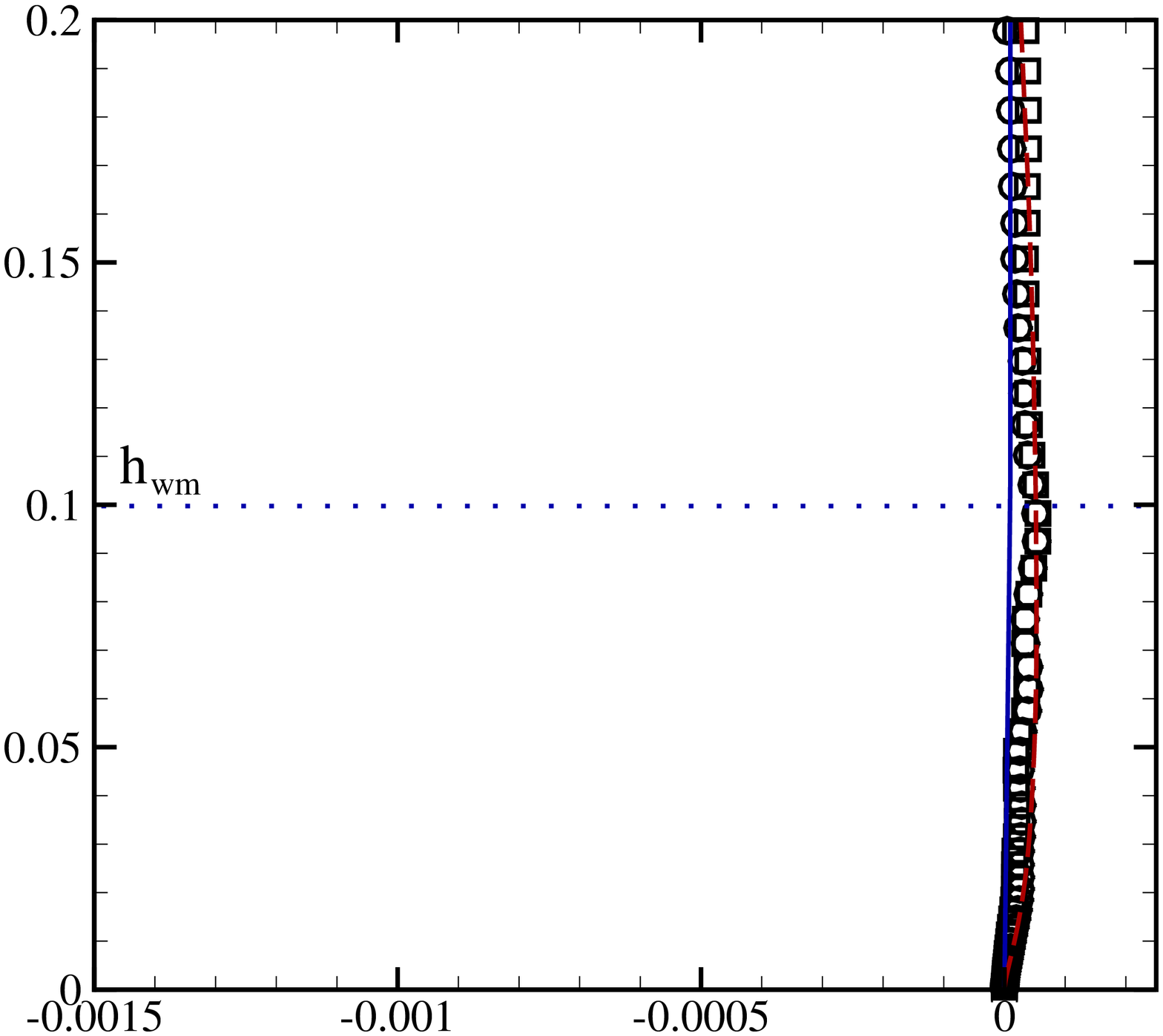}
      \put( -165.0,130.0 ){\makebox(0,0)[r]{\footnotesize$ y / \delta $}}
      \put(  -80.0,  5.0 ){\makebox(0,0)[c]{\  }}
      \vspace{-1em}
    \end{minipage}
    \caption{ \label{fig_uux_param}
      Parametrization of the convective term in terms of the
      pressure-gradient and velocity profile for the APGTBL
      case at stations 1, 3, 5 and 8 using the time-scale $\tau \rightarrow \infty$.
      Approximate $\wt{u}_j \partial_j \wt{u}$ from
      Eq.~(\ref{eqn_uux_parametrized_dpdx}) (\solid),
      Eq.~(\ref{eqn_uux_parametrized_ux}) (\dashed) ,
      exact $\wt{u} \partial_1 \wt{u}$ (\circle), and
      exact $\wt{u}_j \partial_j \wt{u}$ (\square).}
  \end{center}
\end{figure*}

The previous section argued and showed that inclusion of the
pressure-gradient term implies that the convective term must also be
included for consistency reasons.
The convective term, however, includes derivatives in the
wall-parallel directions, which implies that a regular grid with full
connectivity in all directions is needed to solve the wall-model.
While this can be done relatively easily for academic test cases
\cite[cf.][]{balaras:96,wang:02,kawai:10},
it is hard to imagine an implementation in a general-purpose code with
an unstructured grid topology.
Therefore, it is crucial to remove the need for wall-parallel
derivatives in the wall-model.

This can be done by parameterizing the convective term
$\wt{u}_j \partial_j \wt{u}$
in terms of outer
layer quantities, which are available from the LES.
The most straight-forward parametrization stems directly from the
results shown and discussed in Section~\ref{section_consistency}
above.
Since $\wt{u}_j \partial_j \wt{u}$ is essentially balanced by the
pressure-gradient $-\partial_1 \ol{p} / \ol{\rho}$ above the viscous
layer, it follows directly that one can approximate
\beq
\label{eqn_uux_parametrized_dpdx}
\wt{u}_j \partial_j \wt{u} \approx
- \left. \frac{\partial_1 \ol{p}}{\ol{\rho}} \right|_{h_{\rm wm}}
\cdot
\left\{
\begin{array}{ll}
y/y_{\rm pg} &, \ y < y_{\rm pg} \\
1 &, \ y \geq y_{\rm pg} \\
\end{array}
\right.
\eeq
where $y_{\rm pg}$ is the point where viscous effects start damping
the streamwise convective term (akin to the thickness of a Stokes
layer).
For an attached turbulent boundary layer, the value of $y_{\rm pg}$
should be specified in viscous (plus) units for validity across
different Reynolds numbers.
Since the purpose of this study is to enable wall-models to capture
separating flows, we instead set this parameter in viscous pressure-gradient
scaling $y_{\rm pg}^\star = y_{\rm pg} (\rho_w |\partial_1 p| /
\mu_w^2)^{1/3}$; a fixed value of $y_{\rm pg}^\star = 4$ is used
throughout here, with little attempts made at finding an optimal value.

Equation~(\ref{eqn_uux_parametrized_dpdx}) implies that the net effect
of convection and pressure-gradient is zero above $y_{\rm pg}$. Thus
this parametrization predicts that
the logarithmic slope of the mean velocity is independent of the
pressure-gradient, but that the additive intercept constant is not (if
a regular mixing-length eddy viscosity model is used, such as
Eq.~(\ref{eqn_mixinglength})).

A second potential parametrization of the convective term
is to assume that the streamwise component $\wt{u} \partial_1 \wt{u}$
is dominant in the unsteady type of boundary layer flow that occurs in
a wall-model, and then to assume that the vertical shape of the
derivative $\partial_1 \wt{u}$ can be modeled by the shape of the
velocity profile $\wt{u}$ itself.
In other words, to approximate the convective term as
\beq
\label{eqn_uux_parametrized_ux}
\wt{u}_j \partial_j \wt{u} \approx
\left\vert
\frac{\wt{u}}{\left. \wt{u} \right|_{h_{\rm wm}}}
\right\vert^\alpha
\left. \wt{u}_j \partial_j \wt{u} \right|_{h_{\rm wm}}
\,,
\eeq
where the convective term at height $h_{\rm wm}$ is taken from the LES. 
We found that this approach leads to good predictions that depend only weakly 
on the precise values of the free parameter $\alpha$; 
throughout this study $\alpha = 3/2$ is used.

These parameterizations of the convective term are tested
{a priori} on the APGTBL case
in Figure~\ref{fig_uux_param}.
First, note that the infinite time-filtering for this test case
implies that $\wt{w} \partial_3 \wt{u} = 0$.
Secondly, while the $\wt{v} \partial_2 \wt{u}$ term is not
insignificant, it is small compared to the streamwise component
$\wt{u} \partial_1 \wt{u}$. 
The parametrization in terms of velocity, Eq.~(\ref{eqn_uux_parametrized_ux}), gives a
very reasonable agreement with the wall-resolved LES, while the 
parametrization in terms of pressure gradient,
Eq.~(\ref{eqn_uux_parametrized_dpdx}),
only captures the gross features.
However, as will be seen below, this is not the complete story.

\section*{\uppercase{{A priori} validation}}

To assess the two proposed parameterizations of the convective term, a
different type of {a priori} test is performed.
Data from a height $h_{\rm wm}$ above the wall is taken from the
wall-resolved reference LES databases and used as the top boundary
condition for the wall-model equations; these equations are then
solved, and the wall stress is extracted and compared to the actual
wall stress in the reference LES databases.

The wall-model is defined by Eq.~(\ref{eqn_timefiltered_momentum}),
with the unresolved convective term (the last term) modeled using
an eddy-viscosity hypothesis and Eq.~(\ref{eqn_mixinglength}), and
where the sum of the temporal and convective terms (the first two
terms) is modeled using either (\ref{eqn_uux_parametrized_dpdx}) or
(\ref{eqn_uux_parametrized_ux}). In the present study, 
the simple mixing-length model
\beq
\label{eqn_mixinglength}
\mu_{t,{\rm wm}} = \kappa \ol{\rho} y \sqrt{\frac{\wt{\tau}_w}{\ol{\rho}}}
\left[1-\exp\left(\frac{-y^+}{A^+}\right)\right]^2 \,,
\eeq
with $\kappa=0.41$ and $A^+=17$ is used.

A finite volume approach is used to discretize the equations, and
convergence is achieved using a Newton-type iterative procedure.
Identical formulations are used for both databases with identical parameters,
and compared with an equilibrium wall-model as well as the
non-equilibrium wall-model of \cite{duprat:11}.

\begin{figure*}
  \begin{center}
    \begin{minipage}[t]{0.49\textwidth}
      \includegraphics[width=0.97\textwidth,clip=]{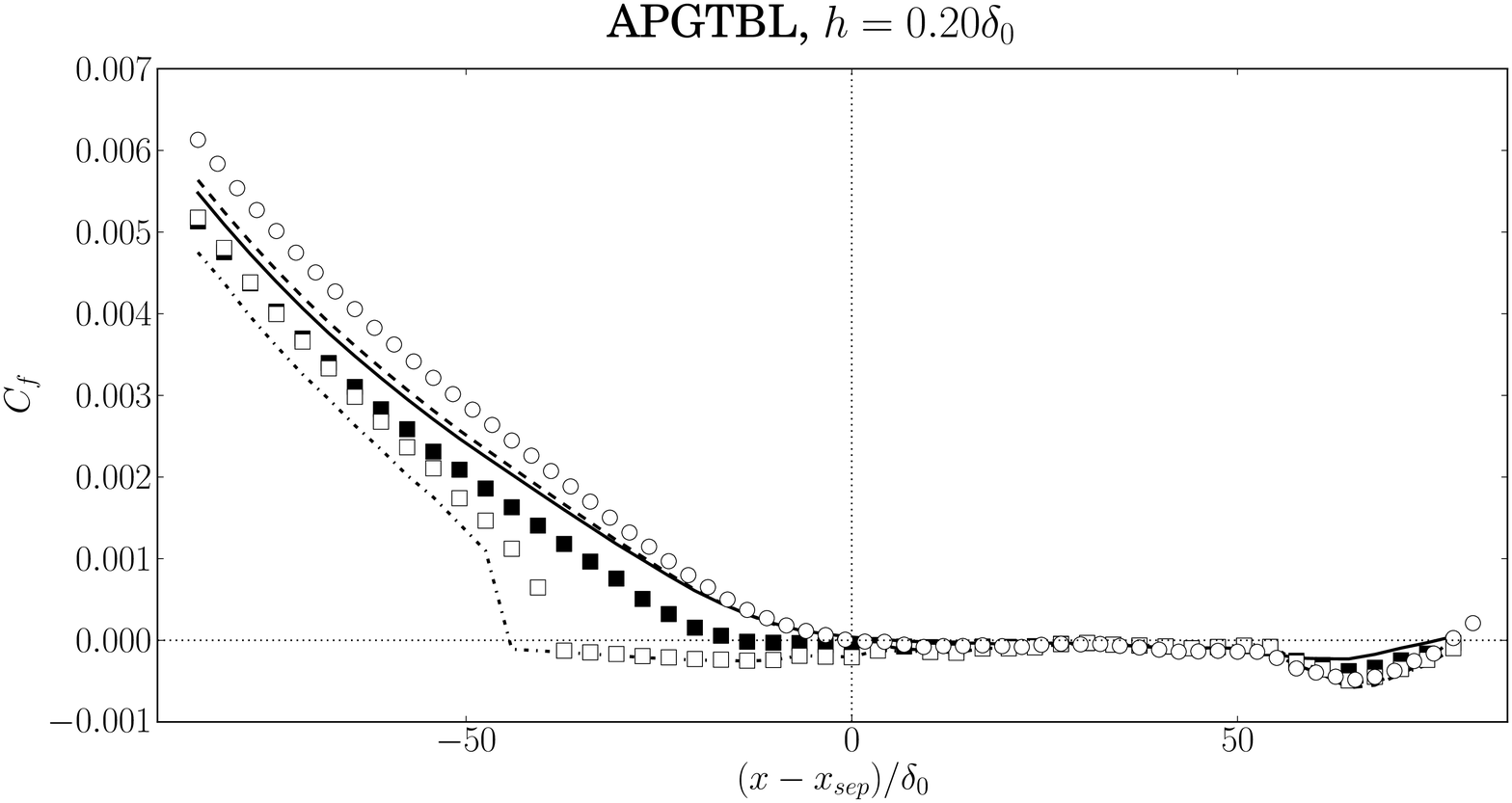}
    \vspace{-1em}
    \end{minipage}
    \hfill
    \begin{minipage}[t]{0.49\textwidth}
     \includegraphics[width=0.97\textwidth,clip=]{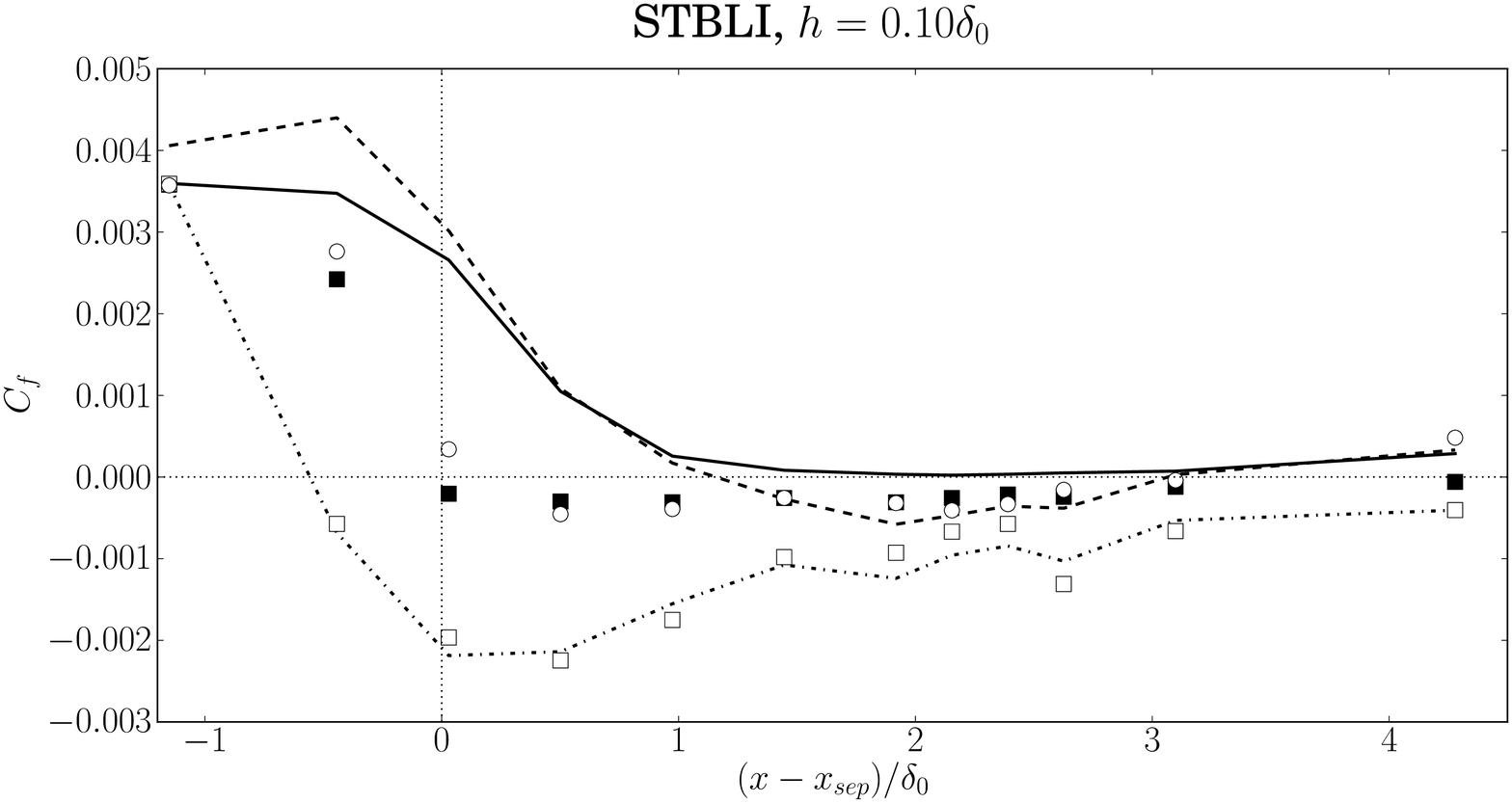}
   \end{minipage}
    \caption[]{ \label{fig_cf}
      A-priori study: Friction coefficient obtained using various non-equilibrium formulations and compared to existing models.
      Exact from reference LES (\circle),
      equilibrium model (\solid),
      model of \citet{duprat:11} (\dashed),
      adding $\partial_1 p$ by itself (\chndot),
      adding $\partial_1 p$ + Eq.~(\ref{eqn_uux_parametrized_dpdx}) with $(y_{pg}^*=4)$ (\solidsquare),
      adding $\partial_1 p$ + Eq.~(\ref{eqn_uux_parametrized_ux}) (\square).
    }
  \end{center}
\end{figure*}

The results are shown in Figure~\ref{fig_cf}.
In the APGTBL case, all of the different
wall-models under-predict the skin friction; this could be due to
the low Reynolds number in this case ($Re_{\tau}\approx 400$ at
station 1), for which the wall-model parameters are not optimal.

If the pressure-gradient is included without any additional modeling
\cite[as done by][]{hoffmann:95, wang:02, catalano:03, chen:13},
the skin friction is further under-predicted, and separation occurs
much too early.

Including both the pressure-gradient and a parametrized convective
term gives a model which satisfies the necessary ``Bernoulli
consistency.''
Despite this, and despite producing impressive {a priori}
agreement of the vertical profiles in 
Fig.~\ref{fig_uux_param},
the parametrization based on the
velocity profile, Eq.~\eqref{eqn_uux_parametrized_ux}, gives disappointing results, hardly better than
without the convective term at all.
In contrast, the parametrization 
based on the pressure-gradient, Eq.~\eqref{eqn_uux_parametrized_dpdx}, gives excellent results for the
STBLI case and reasonable results for the APGTBL case, albeit a bit worse than the
results of the basic equilibrium model.
We also note that the parametrization based on the velocity profile is
more sensitive to numerical convergence issues, while the parametrization
based on the pressure-gradient was found to be 
numerically robust.

The model of \cite{duprat:11} gives disappointing results, essentially
no different from the basic equilibrium model for the APGTBL case and with
severely over-predicted wall shear for the STBLI case.

\section*{\uppercase{Summary}}

Wall-models are essential for enabling the use of large-eddy
simulations on realistic problems at high Reynolds numbers.
The present study is focused on approaches that directly model the
wall shear stress, specifically on filling the gap between models
based on wall-normal ODEs that assume equilibrium and models based on
full PDEs that do not.
Ideas for how to incorporate non-equilibrium effects (most
importantly, strong pressure-gradient effects) in the wall-model while
still solving only wall-normal ODEs are developed and tested using two
reference databases computed using wall-resolved LES: an adverse
pressure-gradient turbulent boundary-layer and a shock/boundary-layer
interaction problem, both of which lead to boundary-layer separation
and re-attachment.

First, it is pointed out that the convective term and the
pressure-gradient term must be treated consistently with each other,
since a non-zero pressure-gradient is almost necessarily associated
with a non-zero convective acceleration; these terms will have
offsetting contributions in most cases.
The bottom line is that these terms should either be retained or
neglected jointly, not independently as done in several prior studies
\cite[e.g.,][]{hoffmann:95,wang:02,catalano:03}.
Similarly, since the temporal and convective terms jointly describe
the acceleration of a fluid particle in its Lagrangian frame, for
consistency these two terms must be treated in the same way as well.

Next, it is argued that a non-equilibrium wall-model in ODE-form
requires that the convective terms be parametrized using LES data from
the top of the wall-modeled layer.
Two forms of this parametrization are proposed: one based on the
pressure-gradient, one based on the velocity profile and the LES
velocity gradient.
When assessed {a priori} using the reference databases, no clear
conclusion is reached: the pressure-based parametrization can capture only
the gross features of the convective term, whereas
the second parametrization based on the velocity profile gives a
very good agreement with the wall-resolved LES data.
However, when used to compute the skin friction in the two test cases, the
model based on the pressure-gradient appears superior:
the predicted skin friction is very close to the reference one for the shock/boundary-layer
interaction case, but slightly under-predicted for the adverse
pressure-gradient case.





\end{document}